\documentclass[useAMS,usenatbib]{mn2e}
\usepackage{mnoverride}
\usepackage[fleqn]{amsmath}
\usepackage{amssymb}
\newcommand{\Msun}{\mathrm{M}_\odot}
\newcommand{\Lsun}{\mathrm{L}_\odot}
\newcommand{\kms}{\mathrm{km~s^{-1}}}

\title[The kinematic signature of massive BBHs]{The kinematic signature of the inspiral phase of massive binary black holes}

\author[Y. Meiron and A. Laor]{Yohai Meiron$^{1}$\thanks{E-mail: ymeiron@pku.edu.cn (YM); laor@physics.technion.ac.il (AL)} and Ari Laor$^{2}$\\
$^{1}$Kavli Institute for Astronomy and Astrophysics at Peking University, Beijing 100871, China\\
$^{2}$Department of Physics, Technion -- Israel Institute of Technology, Haifa 32000, Israel}

\begin{document}

\date{Accepted 2013 May 23. Received 2013 May 22; in original form 2013 February 14}

\pagerange{\pageref{firstpage}--\pageref{lastpage}} \pubyear{2013}

\maketitle

\label{firstpage}

\begin{abstract}
Supermassive black holes are expected to pair as a result of galaxy mergers, and form a bound binary at parsec or sub-parsec scales. These scales are unresolved even in nearby galaxies, and thus detection of non-active black hole binaries must rely on stellar dynamics. Here we show that these systems could be indirectly detected through the trail that the black holes leave as they spiral inwards. We analyze two numerical simulations of inspiralling black holes (equal masses and 10:1 mass ratio) in the stellar environment of a galactic centre. We studied the effect of the binary on the structure of the stellar population, with particular emphasis on projected kinematics and directly measurable moments of the velocity distribution. We present those moments as high-resolution 2D maps. As shown in past scattering experiments, a torus of stars counter-rotating with respect to the black holes exists in scales $\sim 5$ to 10 times larger than the binary separation. While this is seen in the average velocity map in the unequal mass case, it is obscured by a more strongly co-rotating outer region in the equal mass case; however, the inner counter-rotation could still be detected by studying the higher moments of the velocity distribution. Additionally, the maps reveal a dip in velocity dispersion in the inner region, as well as more pronounced signatures in the higher distribution moments. These maps could serve as templates for integral field spectroscopy observations of nearby galactic centres. The discovery of such signatures may help census the population of supermassive black hole binaries and refine signal rate predictions for future space-based low frequency gravitational wave detectors.
\end{abstract}


\section{Introduction}

Supermassive black hole (BHs) are expected to form binary systems as a result of galaxy mergers \citep{BBR}, commonly believed to be the primary mechanism for galaxy growth in $\Lambda$CDM cosmology (e.g. \citealt{Kauffmann1993}; cf. \citealt{Dekel2006}). Prior to the formation of a binary, the two BHs will form an unbound pair, moving independently in the stellar potential, on shrinking orbits due to dynamical friction. In case of a gas-rich merger, accretion onto both BHs could be triggered, forming a dual active galactic nucleus (AGN).

Indeed, there are many examples known of dual (unbound) AGN. \citet{Liu2011} surveyed AGN from the Sloan Digital Sky Survey (SDSS) at $z \sim 0.1$ and found that the fraction of AGN pairs with projected separations between $\sim 5$ and 100~kpc is 3.6 per cent. This frequency is consistent with the galaxy merger rate under some reasonable constraints \citep{Yu2011}.

It becomes increasingly harder to discover a double AGN with decreasing separation, and at the 1 kpc scale only a handful of convincing examples is known \citep{Fu2012}. At even smaller scales, there is the single fiducial example of radio galaxy 0402+379, where the projected separation of 7.3~pc between the two component could only be resolved by the Very Long Baseline Array (VLBA; \citealt{Rodriguez2006}).

Evidence of binary (bound) black holes (BBHs) is virtually nonexistent. Imaging is extremely challenging; 0402+379, in which the BHs are still too widely separated to form a bound pair, could not have been discovered as a dual system even with VLBA, had the separation shrunk to $\sim 1$~pc or below, or had there been too little or no radio emission from one or both BHs. Alternative techniques to search for subparsec binaries have not produced convincing results. The quasar SDSS J153636.22+044127.0 was advanced as a candidate BBH by \citet{Boroson2009} due to its double-peaked broad lines, however it failed to show a velocity shift within $\sim 1$ year and is more likely an unusual double-peaked emitter (\citealt{Lauer2009}; \citealt{Chornock2010}). It has been claimed that variable blazar OJ 287 is sub-parsec BBH system due to an apparent regular 12-year double-peaked outbursts (\citealt{Silanpaa88} and references thereafter), however, \citet{Villforth2010} tested BBH models for this object and found that they could not explain the observations. Similarly, claims that 3C~66b is a BBH \citep{Sudou2003} due to supposed elliptical motion of the compact radio core have been refuted due to the failure to detect the corresponding fluctuations in the pulse-arrival times of a nearby pulsar \citep{Jenet2004}.

The lack of observations of binaries is not surprising when considering the evolution of BBHs. Initially, the two BHs sink independently due to dynamical friction from the scale of the galactic merger (10-100 kpc) down toward the bottom of the stellar potential. The infall time at a given $r$, i.e. $r/\dot{r}$ for a singular isothermal sphere (following \citealt{Just2005}) is:
\begin{align}
t_{\rm df} &\sim 1.4 \times 10^8\ {\rm yr}\notag\\
 &\phantom{\sim} \times
 \left( \frac{\ln \Lambda}{5.23} \right)^{-1}
 \left( \frac{M_\bullet}{10^8\ \Msun} \right)^{-1}
 \left( \frac{\sigma}{200\ \kms} \right)
 \left( \frac{r}{1\ {\rm kpc}} \right)^2
\end{align}
where $M_\bullet$ is the BH mass, $\sigma$ is the velocity dispersion, $r$ is the distance from the centre and $\Lambda = M(r)/M_\bullet = 2\sigma^2r/GM_\bullet$ is the modified Coulomb logarithm. This timescale gets shorter as $r$ decreases, which means that there is a decreasing probability to find BBHs with decreasing separation. Once almost all stars on low angular momentum orbits have interacted with the BBH (lost-cone emptying), its orbital evolution reaches a bottleneck (the `final parsec problem') and the above expression is no longer valid. However, \citet{Yu2002} showed that surviving BBHs even in nearby galaxies have semi-major axes, which are just within the {\it Hubble Space Telescope} resolution of 0.1 arcsec or below. The BBH spends most of its life in either the galactic scale (where it is not a proper binary), or the bottleneck stage (where it is most likely unresolved). Thus, very compact, but still resolved, BBHs, should be very rare. This is the motivation for our work: to find an evidence for a hard BBH on larger angular scales.

The subsequent evolution of BBHs beyond this bottleneck and possible coalescence via gravitational radiation, is a separate topic and not touched upon in this work. We note that the general consensus is that BBHs get through the bottleneck in less than a Hubble time either via centrophilic orbits in triaxial stellar systems (e.g. \citealt{Preto2011}; \citealt{Khan2011}; \citealt{Khan2013}), presence of massive perturbers \citep{Perets2008} or gas in gas-rich mergers (e.g. \citealt{Escala2005}; \citealt{Dotti2007}; \citealt{Cuadra2009}).

In \citet{Meiron2010} we proposed that BBHs might be detectable through the significantly anisotropic stellar velocity distribution of nearby stars, on larger scales than the binary separation. We calculated the line-of-sight velocity distributions (LOSVDs) of stable orbits near a stalled BBH (circular orbit with constant separation) by solving the restricted three-body problem for a BBH embedded in a bulge potential. The LOSVD can be directly measured from stellar absorption lines, and be used to diagnose orbital structure anisotropies (e.g \citealt{Dejonghe1987}; \citealt{Bender1990};  \citealt{Gerhard1993}; \citealt{vdMarel1993}). Those anisotropies are mainly expressed as the deviation of the LOSVD from a normal distribution; and to characterize these deviations one often uses Gauss--Hermite (GH) moments (\citealt{Gerhard1993}; \citealt{vdMarel1993}). Our results were presented as high-resolution maps of the LOSVD GH moments, which could be used as templates for Integral Field Spectroscopy BBH searches.

In this work, we improve on that by using more realistic models of a stellar system with BHs from \citet{Meiron2012}, where by imposing conservation of total energy and angular momentum in scattering experiments, we could follow the BHs' orbital decay and the response of the stellar population. While our earlier work was for a steady-state hard binary, the formation of the hard binary will also affect the background stellar distribution well before the BBH becomes hard. In particular, from the stage the dual BH becomes bound, and during its inspiral until the hard binary phase. The purpose of this study is to find the signature induced also during the inspiral phase. This signature is expected to be set on significantly larger scales, and may thus may be observable in nearby galaxies.

In Section \ref{sec:new-results-data} we describe the simulation from which are results were derived. In Section \ref{sec:KinMaps} we present the main result of our new analysis of stellar kinematics. In Section \ref{sec:photometry} we show the brightness profiles of our models as inspired by observations, and analyse the mass deficit that developed in the system due to the presence of the BBH. Finally, in Section \ref{sec:discussion} we discuss the origin and implication of our results and briefly summarize in Section \ref{sec:conclusions}.

\section{Simulation data}\label{sec:new-results-data}

In this work, we analysed two simulations that have already been published in \citet{Meiron2012}. The simulations are of an equal mass and 10:1 mass ratio BBHs. In both there were $5\times10^6$ stars, in an initially isothermal sphere, with a cutoff at $R_{\rm max}=200$ simulation length units as explained below, and the initial semi-major axis of the binary was $a_0=60$ units (almost one order of magnitude more than the primary BH's radius of influence). The two BHs were initially on a circular orbit in the equal mass case, while the initial eccentricity\footnote{Eccentricity was defined geometrically, as the difference between apo- and pericentre divided by their sum, had the BHs been test particles travelling in the stellar potential, with no losses.} was $e_0=0.2$ in the unequal mass case. The simulations were performed using the conservation-based method presented in that paper, where the forces on the BHs are derived from the stars' change in energy and angular momentum, under some constraints.

The units of measurement in the results presented in this work are the same as in \citet{Meiron2012}. Mass is measured in units of the primary BH's mass $M_\bullet$, the velocity unit is four times the velocity dispersion $\sigma$ and $G = 1$ (where $G$ is the gravitational constant). These units were set so that the hard binary separation, equals two length units in the case of an equal mass binary. The scaling to physical units thus has two parameters, $M_\bullet$ and $\sigma$. However, only one parameter is required if the $M$--$\sigma$ relation (e.g. \citealt{m-sigma}) is used. The units of length, time and velocity and their scalings are
\begin{align}
[\mathrm{L}] &= {\textstyle\frac{1}{16}} GM_\bullet\sigma^{-2} = 0.77\ M_8^{0.53} \ \mathrm{pc},\\
[\mathrm{T}] &= {\textstyle\frac{1}{64}} GM_\bullet\sigma^{-3} = 1\,000\ M_8^{0.29} \ \mathrm{yr},\\
[\mathrm{V}] &= 4\sigma = 750\ M_8^{0.24} \ \mathrm{km\ s^{-1}},
\end{align}
where $M_8$ is the physical mass of the primary BH in units of $10^8\ \Msun$. In Section \ref{sec:photometry} we used observational units, as described there.

Briefly describing the evolution of the BBH in these simulations, as more fully discussed in \citet{Meiron2012}: in the unequal mass case, the two BHs form a bound pair at $t\approx 3\,900$ with $a \approx 0.8$ and $e \approx 0.15$. We superimposed ten snapshots $\sim 120$ time steps apart and starting from $t=10\,000$, by which time the orbital elements were $a=0.24$ and $e=0.52$. Eccentricity is increasing at a nearly constant rate of $\dot{e}=7\times10^{-5}$. The time interval between snapshots is large enough compared to the orbital time, so that the the stellar positions are not correlated, which smooths the final image.

In the equal mass case, the two BHs form a bound pair at $t\approx 1\,600$ with $a \approx 1.3$ (in physical units, $1\ \mathrm{pc} \times M_8^{0.53}$) and $e \approx 0.01$. We superimposed six snapshots, $1\,000$ time steps apart and starting from $t=40\,000$, by which time the orbital elements were $a=0.68$ and $e=0.009$. In this simulation, eccentricity dropped to $\approx 0.004$ and climbed back up slightly. The reason that in this case the snapshots are spaced much further apart in time, is that the system evolution is slow as compared to the unequal mass case. This allowed us to use existing snapshots (for the unequal mass case, the simulation had to be rerun for a short period in order to generate more tightly spaced snapshots). To further increase the statistics, we performed angular averaging of every snapshot as justified and explained in Section \ref{sec:KinMaps}. This applies to both equal and unequal mass cases.

\section{Kinematic maps}\label{sec:KinMaps}

\begin{figure*}
\includegraphics{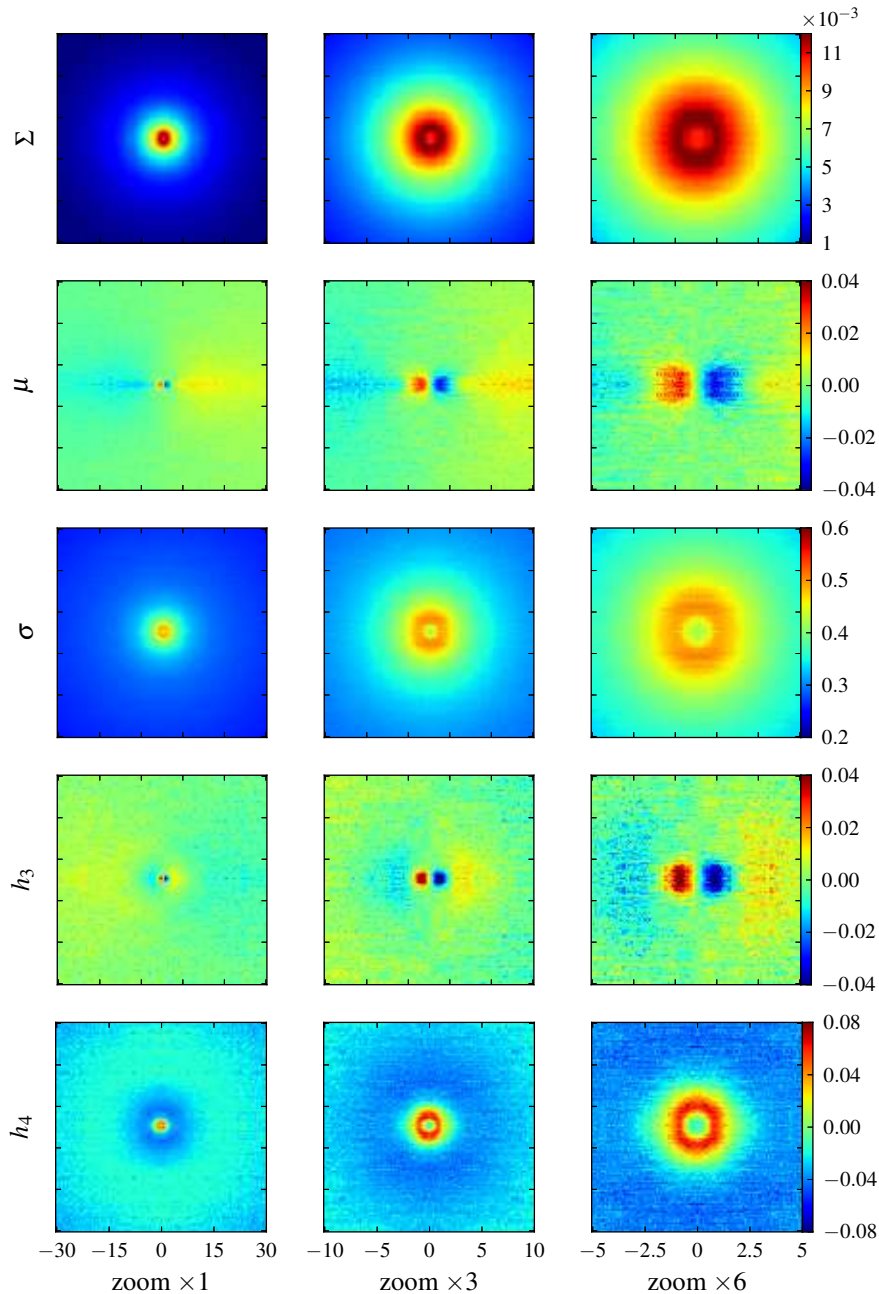}
\caption{Edge-on projected kinematics and density of a 10:1 mass ratio inspiral. The three panels in each row are different zoom levels; the upper row is projected density (mass per unit area in simulation units) and below are moments of the line-of-sight velocity distribution (the two bottom rows are Gauss-Hermite moments). The semi-major axis of the binary when this snapshot is taken is 0.24 and the eccentricity $e = 0.52$. Fig. \ref{fig:slit} shows a view of these projections through a slit.}
\label{fig:KMaps-unequal}
\end{figure*}

\begin{figure*}
\includegraphics{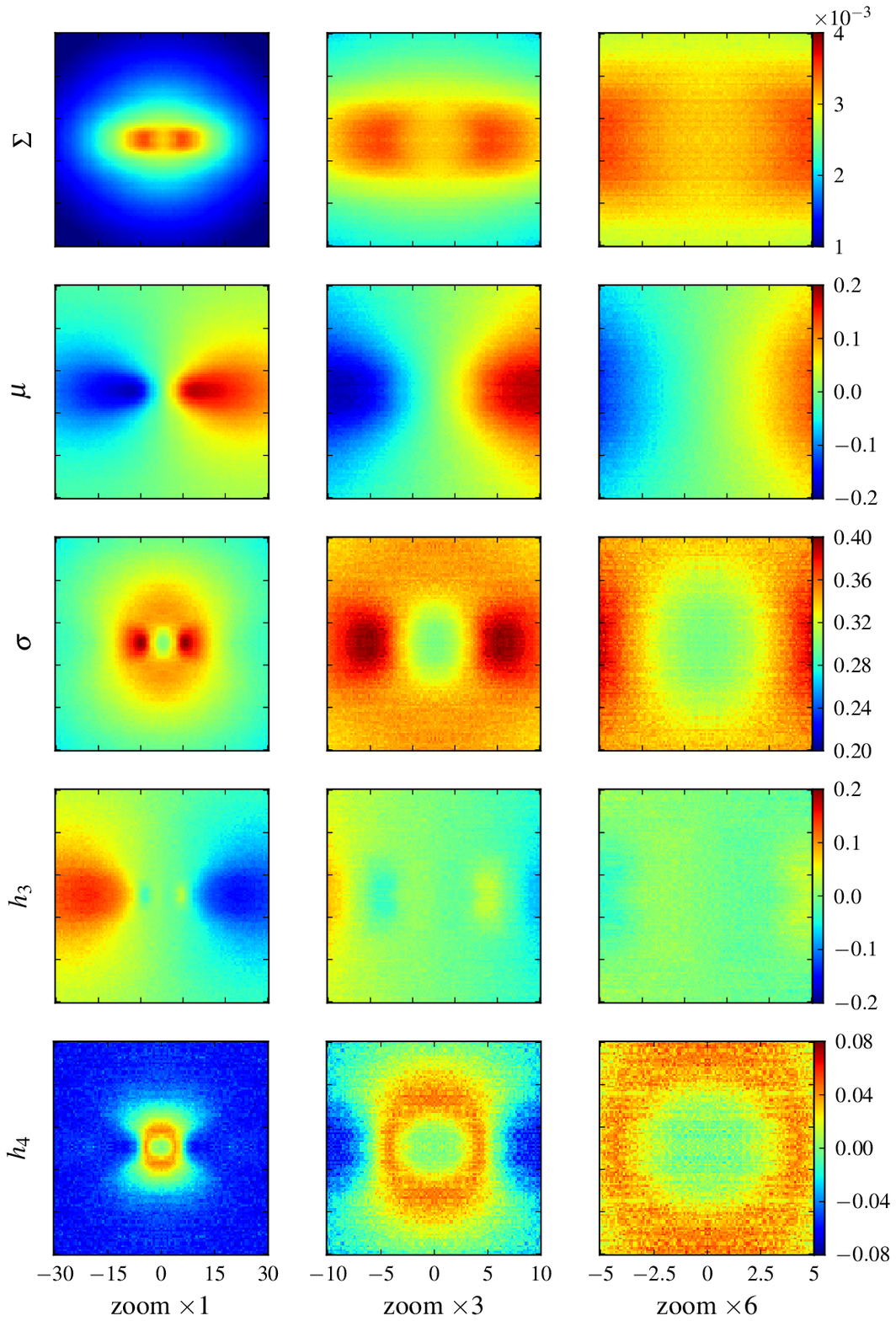}
\caption{Same as Fig. \ref{fig:KMaps-unequal}, but for the equal mass case.}
\label{fig:KMaps-equal}
\end{figure*}

\begin{figure*}
\includegraphics{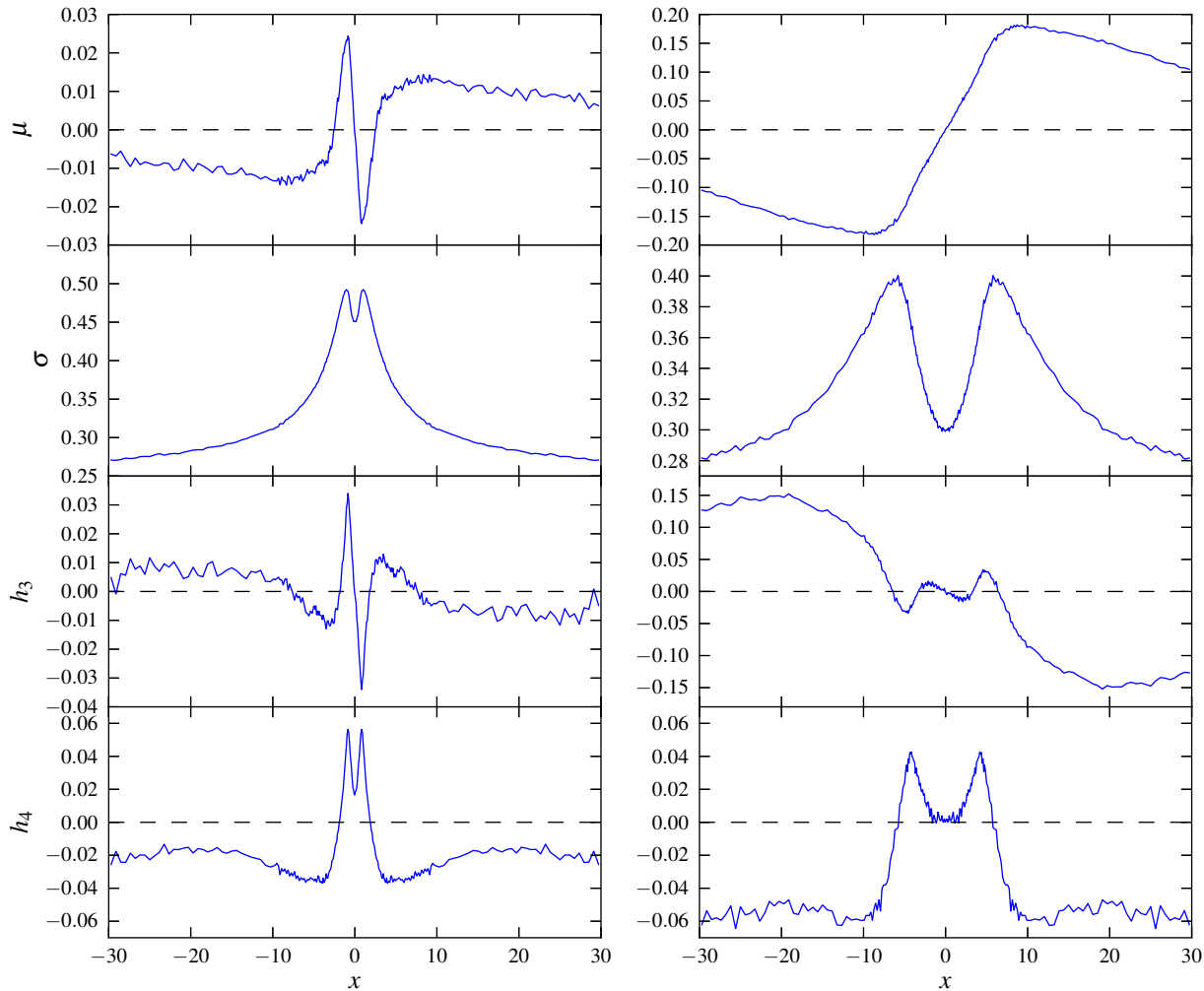}
\caption{The line of sight velocity distribution moments from Figs. \ref{fig:KMaps-unequal} (left) and \ref{fig:KMaps-equal} (right), viewed through a slit along the horizontal axis with a width of 3.2 length units.}
\label{fig:slit}
\end{figure*}

\begin{figure}
\includegraphics[width=1\columnwidth]{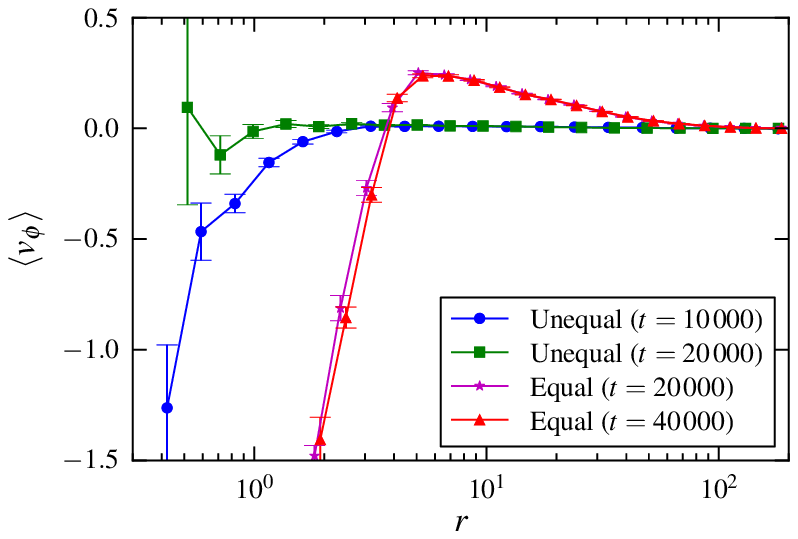}
\caption{Average tangential velocity in different radius bins for the unequal mass case: blue circles and green squares for the early and late snapshots, respectively; and the equal mass case: magenta stars and red triangles for the early and late snapshots, respectively (times indicated in the legend). Note that $\langle v_\phi \rangle$ is not the same as the average projected velocity $\mu$, but rather represents the internal orbital structure. The errorbars are calculated from the velocity dispersion in each radius bin, divided by the square root of the number stars per bin.}
\label{fig:internal}
\end{figure}

Figs. \ref{fig:KMaps-unequal} and \ref{fig:KMaps-equal} show the projected kinematics for the unequal and equal mass cases, respectively. They have five rows showing (from top down) surface density, average velocity, velocity dispersion, third and fourth Gauss-Hermite moments of the line-of-sight velocity distribution (LOSVD), for a galaxy observed from the orbital plane of BBH (i.e. {\it edge-on}). The quantities are given in simulation units (i.e. surface density is in mass per unit length square); the GH moments are dimensionless and defined below. The three panels at each row are different zoom levels as indicated by the axis labels. The vertical axis is the $z$-axis or the direction of the BHs' angular momentum. The horizontal axis is really a combination of $x$ and $y$: to increase the statistics further (beyond superimposing different snapshots as described above), each snapshot was superimposed with $X$ copies of itself, each one rotated by a different angle along the $z$-axis (we refer to this as `azimuthal averaging'); $X$ is typically 10--100, and is set so all maps have roughly the same quality (i.e. number of effective particles).

The assumption of azimuthal symmetry of the system is very good when the BHs are tightly bound, because their orbital period is much shorter than that of surrounding stars. However, there is a preferred axis on the orbital ($xy$-) plane (unless the orbit happens to be circular), this is the eccentricity vector, which precesses in a much slower rate compared to the orbital periods of stars in the vicinity. Since the simulation code does not include general-relativistic corrections, the precession of the eccentricity vector is purely Newtonian. Azimuthally-dependent features, while possible, were not seen in this setup; this might be due to poor signal to noise ratio of a single snapshot without the azimuthal averaging. Dependence on the polar angle $\theta$ is still very much expected, so spherical symmetry could not be assumed. The horizontal axis will be referred to as the $x$-axis for convenience. The aspect ratio of the maps is 1:1, for convenience, just the $x$-axis is labelled.

For each pixel $(x,z)$ we obtained the LOSVD from a histogram of the $y$-direction velocities of all stars which were mapped to that pixel. We added to the LOSVD a correction term representing the stars that lie outside the simulation volume (`bulge correction'): since the simulation had a cutoff at $R=200$ and contained total stellar mass of $\approx 25 M_\bullet$, we added a Gaussian component with weight corresponding to the column of stars between $R_{\rm max}=200$ and $R=5\,600$, so that the effective mass of the entire bulge would be $700 M_\bullet$ (corresponding to the bulge to BH mass ratio found by \citealt{Haring2004}); this correction was typically extremely small since the stars are highly concentrated in a power-law profile. We then considered a generic line profile of the form
\begin{align}
\mathcal{L}(v) = \frac{\gamma}{\sqrt{2 \pi}\sigma}e^{w^2/2}
\left[ 1 + \sum_{n=3}^N h_n H_n(w) \right],\label{eq:formula}
\end{align}
where $w=(v-\mu)/\sigma$ is the normalized velocity parameter and $H_n(w)$ is the Hermite polynomial of the $n$-th degree. We then found the simultaneous best fitting $\gamma$, $\mu$, $\sigma$, and $h_n$ ($n\geq 3$) using the least-squares method. The first three quantities give the best fitting Gaussian while the GH moments characterize deviation from the zeroth-order Gaussian profile. This procedure is consistent with \citet{vdMarel1993} and assures completeness of the above series (i.e. $h_1=h_2=0$).

Fig. \ref{fig:slit} shows a view of these projections through a slit along the $x$-axis. In both unequal and equal mass cases, there is an outer region co-rotating with the BHs (expressed as positive projected velocity on the right side of the $x$-axis and negative on the left). This region is rotating because of the angular momentum transferred to the stars from the inspiralling BHs. It was not seen in \citealt{Meiron2010}, where there was no inspiral. In the unequal mass case, an inner counter-rotating region exists, but this region is not seen in the equal mass case average velocity map. In \citet{Meiron2010}, counter-rotation appeared in both unequal and equal mass models; as shown there, it is caused by the asymmetry of the loss-cone, which is due to the preferential scattering of co-rotating orbits that leads to an excess of counter-rotating orbits on scales of 5--10 times the binary separation; namely, counter-rotating orbits are more stable.

A counter-rotating region does in fact exist in the equal mass case as well. This is clearly seen in Fig. \ref{fig:internal}, which shows the internal, rather than projected, kinematics. The data points are the average tangential velocity $v_\phi \equiv (xv_y-yv_x)/\sqrt{x^2+y^2}$ in logarithmic radius bins of the unequal and equal mass case snapshots discussed above (blue circles and red triangles, respectively); additionally, green squares represent a late-time snapshot of the unequal mass simulation and magenta stars represent an early snapshot of the equal mass case, both at $t=20\,000$. In the equal mass case, the co-rotation of the outer region dominates the projection, and the average projected velocity $\mu$ does not show sign of this structure. However, the excess of fast counter-rotating stars at small scales still leaves a mark on the $h_3$ map. In the unequal mass case, as shown, the counter-rotation significantly weakens at late times (this is further discussed in Section \ref{sec:discussion}); in the equal mass case the velocity curve is very stable on that timescale.

Another feature of this system is the dip in velocity dispersion (LOSVD width) in the centre (also in agreement with \citealt{Meiron2010}). This feature appears in both models and is due to the evacuation of stars from the immediate vicinity of the BBH; thus, most stars that contribute are from farther out from the centre, and thus have smaller velocity dispersion. This effect is much more prominent in the equal mass case since the evacuated region is larger. Also in the equal mass case there is an elongated shape in the $\sigma$ map, with two peaks roughly 6 length units from the centre. Those two peaks in $\sigma$ also correspond to peaks in the projected density. This feature {\it does not} represent clustering around the two BHs, as they are much closer to the centre than that.

\begin{figure*}
\includegraphics{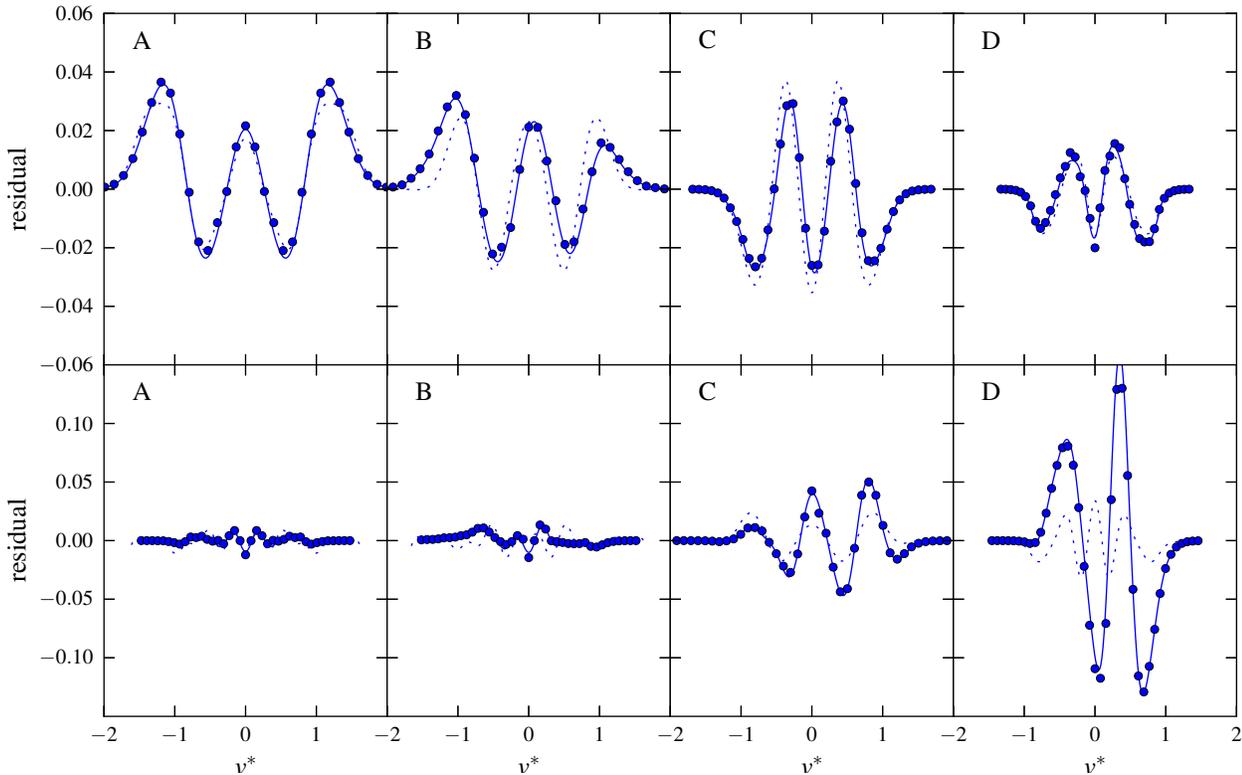}
\caption{Sample velocity distributions from selected lines of sight or `windows', labelled A to D (the top panels correspond to the unequal mass case, and the bottom to the equal mass case). The plots are all residuals: the difference between the LOSVD and the best fitting Gaussian (the distribution is normalized such that it peaks at 1). The circles are the normalized histogram data of the edge-on projected velocities, the solid line is the best fitting model (equation \ref{eq:formula}) up to $n=20$. The dashed line is the same for a face-on view (no histogram data is shown). The windows are square apertures centred on the $x$-with some offset, as described in Tab. \ref{tab:fit}, which also gives the best fitting parameters up to $n=4$ for those lines. Note that a portion of the LOSVD shown here is $10\sigma$ wide.}
\label{fig:LOSVD}
\end{figure*}

A positive (negative) $h_3$ value means that the distribution has a more pronounced right (left) wing and a depression on the opposite wing, it is thus similar to skewness. A positive (negative) $h_4$ value means a symmetric excess (deficiency) on both wing and at the peak. The next odd and even GH moments represent higher order anti-symmetric and symmetric deviations, respectively, and could thus be used to refine the LOSVD characterization; however, since those deviations are increasingly smaller, it might not be very helpful to show their maps. In Fig. \ref{fig:LOSVD} we show the LOSVD in different `windows' or rectangular apertures. Since it is very difficult to see the deviations from a normal distribution in most cases, the figure shows the data minus the best fitting Gaussian, that is, the residual. The data is fitted with a Gauss-Hermite series up to $n=20$, and the first two nontrivial terms appear in Tab. \ref{tab:fit}. Fig. \ref{fig:LOSVD} also shows as a dashed line the LOSVD as seen {\it face-on}, i.e. from above the BH's orbital plane; those LOSVDs are naturally symmetric (due to up-down symmetry of the system) but show, for example, a similar dip in $\sigma$ as in the edge-on case.

\begin{table}
\caption{Best fitting parameter to the velocity distributions shown in Fig. \ref{fig:LOSVD}. The leftmost column is the letter designating the window (panel in Fig. \ref{fig:LOSVD}); $r_{\rm cen}$ is the offset of the centre of the windows; $w$ is its width (the windows are square apertures); $\mu$, $\sigma$, $h_3$ and $h_4$ are the best fitting parameters for the edge-on data (note that for display convenience, some of the parameters are multiplied by 100); the two rightmost columns are the best fitting parameters for the face-on views (note that $\mu$ and $h_3$ are zero due to symmetry).}
\label{tab:fit}
\begin{tabular}{cccrrrrrr}
\hline
\hline
 & $r_{\rm cen}$ & $w$ & $\mu$ & $\sigma$ & $h_3$ & $h_4$ & $\sigma^*$ & $h^*_4$\\
 & & & $\times 100$ & & $\times 100$ & $\times 100$ & & $\times 100$\\
\hline
\multicolumn{9}{l}{Unqual mass case}\\
\hline
A & 0  & 2  & 0    & 0.47 & 0     & 4.1  & 0.47 & 3.7\\
B & 1  & 2  & -1.6 & 0.48 & -1.4  & 2.7  & 0.49 & 2.4\\
C & 5  & 2  & 1.0  & 0.37 & 0.6   & -3.5 & 0.37 & -4.3\\
D & 20 & 10 & 0.9  & 0.29 & -0.7  & -2.0 & 0.28 & -1.9\\
\hline
\multicolumn{9}{l}{Equal mass case}\\
\hline
A & 0  & 2  & 0    & 0.30 & 0     & 0    & 0.33 & -0.4\\
B & 1  & 2  & 2.6  & 0.31 & -0.8  & 0.2  & 0.34 & 0\\
C & 5  & 2  & 13.9 & 0.39 & 2.1   & 1.6  & 0.37 & 2.4\\
D & 20 & 10 & 14.9 & 0.31 & -13.0 & -2.3 & 0.30 & -0.9\\
\hline
\end{tabular}
\end{table}

\section{Photometry}\label{sec:photometry}

\begin{figure}
\includegraphics[width=1\columnwidth]{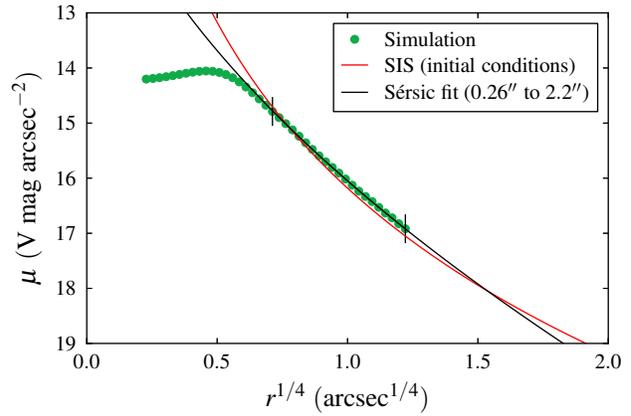}
\caption[Brightness profile (unequal masses)]{`Photometry' of the projected density shown in Fig. \ref{fig:KMaps-unequal} (top row). While the simulation volume is cut off at $\sim 11^{\prime\prime}$, in this figure we added light from the analytic continuation of the stellar distribution up to $\sim 300^{\prime\prime}$ (green circles), so that the total mass of the spheroidal component is 700 times the mass of the primary black hole. The black line is the best fitting S\'ersic model. The solid red line is a singular isothermal sphere (SIS), corresponding to the initial surface brightness. Note the small and unusual `hump'.}
\label{fig:photometry-unequal}
\end{figure}

\begin{figure}
\includegraphics[width=1\columnwidth]{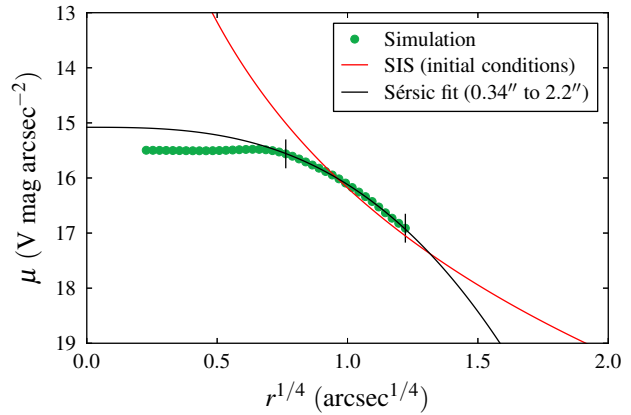}
\caption[Brightness profile (equal masses)]{Same as Fig. \ref{fig:photometry-unequal}, but for the equal mass case.}
\label{fig:photometry-equal}
\end{figure}

Figs. \ref{fig:photometry-unequal} and \ref{fig:photometry-equal} show the brightness profiles of the two models. These two figures are inspired by observations (e.g. \citealt{Kormendy2009}) so we used different units, which are more suited to observers. We assumed a galaxy 3 Mpc away, and with a mass to light ratio $\Upsilon = 8\ \Msun/\Lsun$. The angular length unit with its calibration is
\begin{align}
\mathrm{arcsec} \times M_8^{0.53} \left( \frac{d}{3\ \mathrm{pc}} \right)^{-1},
\end{align}
where $d$ is the distance from the observer to the galaxy. Note however that of the horizontal axis of Figs. \ref{fig:photometry-unequal} and \ref{fig:photometry-equal} shows $r^{1/4}$ rather than $r$. The vertical axis shows the surface brightness $\mu$ (not to be confused with the average projected velocity), and is in the following logarithmic units:
\begin{align}
\mathrm{mag\ arcsec^{-2}} + 2.5 \log_{10}\Upsilon_8 - 5.15\log_{10} M_8,
\end{align}
where $\Upsilon_8$ is the mass to light ratio in units of $8\ \Msun/\Lsun$. Note that this unit is distance-independent and inverted (i.e. higher brightness is lower magnitude). The simulation data (shown as green circles) is the same as shown in the top panels of kinematic maps (Figs. \ref{fig:KMaps-unequal} and \ref{fig:KMaps-equal}) but as a function of projected radius alone (averaged over an annulus ignoring oblateness in the isophotes; in the equal mass case the oblateness seems to be significant); the solid black line is the best fitting S\'ersic model to the data between the points indicated in the legend; the solid red line is the brightness profile of a singular isothermal sphere (SIS), corresponding to the initial conditions.

In both cases, the S\'ersic model fits the data well within the selected limits; the fit breaks at small $r$, as in real galaxies. From the difference between the extrapolation of the S\'ersic function to the centre and the data, we calculated the mass deficit below. Note that the equal mass case is dimmer at the centre, as stars are more effective evacuated from the central region.

The data in the photometric figures as well as the top panels of Figs. \ref{fig:KMaps-unequal} and \ref{fig:KMaps-equal} are corrected for the light of the rest of the spherical component, an isothermal sphere with mass of $700M_\bullet$ (as explained in Section \ref{sec:KinMaps}; the simulated part is just the central $\sim 25M_\bullet$). Dust obscuration was not accounted for, as it is highly variable. The surface density $\Sigma$ is simply the integral of the 3D density $\rho(r)$ along the line-of-sight. For $r<R_{\rm max}$ (note that $r$ is the 3D-radius here rather than a projection) this is calculated from the simulation data by counting the number of particles that fall in each pixel (basically a 2D histogram) after all the superpositions described above and dividing by the number of superimposed snapshots; let us refer to this quantity as $\Sigma_0$. In order to correct for the mass column outside the simulation volume, one needs to add to $\Sigma_0$ the integration of the spatial density at $R_{\rm max}<r$, which is done analytically or numerically depending on $\rho(r)$. In the case of an isothermal sphere, there is a simple solution:
\begin{align}
\Sigma' &= \frac{1}{r_\bot} \left[ \arctan\left(\sqrt{\frac{R_2^2-r_\bot^2}{r_\bot^2}}\right) - \arctan\left(\sqrt{\frac{R_1^2-r_\bot^2}{r_\bot^2}}\right)  \right]\notag\\
 &\phantom{=} \times 1.99\times 10^{-2} \ \mathrm{\frac{[M]}{[L]^2}}\label{eq:light-correction}
\end{align}
where $R_1 \equiv R_{\rm max}$ is the cutoff of the simulation and $R_2$ is the cutoff of the real galaxy (an isothermal sphere has to have a cutoff or its mass diverges; the surface brightness however does not diverge at any point even if the model has infinite mass); $r_\bot=\sqrt{x^2+z^2}$ is the projected distance from the centre. The above quantity is a function of ${r_\bot}$ and $\Sigma_0$ is a function of $x$ and $z$; the sum of the two is the corrected projected mass $\Sigma$ as shown in Figs. \ref{fig:KMaps-unequal} and \ref{fig:KMaps-equal}.

The mass deficit (i.e. the mass associated with the difference between the interpolation of the S\'ersic profile to $r=0$ and the observation) in the unequal mass case is just $0.1M_\bullet$; however, the actual mass scattered in the simulation was $0.55M_\bullet$ (in agreement with \citealt{Merritt2006}). In the equal mass case, the mass deficit is just $0.03M_\bullet$, with the actual mass scattered being $2.7M_\bullet$; this discrepancy is due to the fact that the initial stellar model (dashed red line if Figs. \ref{fig:photometry-unequal} and \ref{fig:photometry-equal}) is significantly more luminous than the S\'ersic fit. 

\section{Discussion}\label{sec:discussion}

We showed that a reverse trend in rotation direction on small scales, as well as a dip in the velocity dispersion near the centre (instead of the normal peak), could be signatures of a BBH in the system. We also recreated a result known from analytical models and $N$-body simulations (e.g. \citealt{Merritt2006}), that a BBH creates a light deficiency at the centre of the galaxy. Such a deficit is seen and seems to be correlated with BH mass \citep{Milos2002, Graham2004, Ferrarese2006, KormendyBender09}.

We were able to obtain those result mainly due to the large number density in our simulation, which enabled us to get good statistics of stellar velocities. The number of stars within the primary BH's sphere of influence is a very simple and model-independent indicator of number density; it is $NM_\bullet/M_{\rm tot}$, where $N$ is the total number of stars and $M_{\rm tot}$ is the total stellar mass. For comparison, while this number is $\sim 2\times 10^5$ in our models, it is only $\sim 2.6\times 10^3$ in the simulations by \citet{Khan2012}. Cf. \citet{Iwasawa2011, Sesana2011} who like us had $\gtrsim 10^5$, however both studies did not explore the full inspiral process but rather began with a secondary BH embedded in the sphere of influence of the primary. While superposition of snapshots and azimuthal averaging help increase the statistics and make unique kinematic features more prominent, it is likely that they would not be helpful below some threshold of number density. More importantly, low number density could adversely affect the BBH evolution: while \citet{Berczik2006} thoroughly explored the effect of $N$ on the hardening rate, the effect on eccentricity growth is not yet understood. Thus, `ultra-dense' simulations are required to understand the system at small scales, these could be achieved by increasing $N$ or by reducing $M_{\rm tot}$, or both.

The high number density in our models was achieved thanks to the conservation-based method \citep{Meiron2012}, which is a powerful tool for exploring problem where a few very massive bodies are surrounded by a sea of lighter particles. Using only a desktop computer, we were able to follow the evolution of a BBH from wide separations down to sub-parsec scales, using as many as $5 \times 10^6$ particles in a simulation. The results were that the lifetimes of these systems, especially when the binary components have unequal masses, could be very short due to rapid increase of eccentricity. The lifetime is possibly even shorter if the stellar distribution is triaxial or rotating. This calculation improved on $N$-body simulation by reducing statistical fluctuations and having no spurious relaxation (and thus no loss-cone refilling).

The lifetime of the system is related, of course, to the lifetime of the kinematic signature. Once kinematic features arise, relaxation will tend to homogenize the system and erase them, unless the generating mechanism is still present to preserve them. In our case, the kinematic signature is due to the unique shape of the loss-cone of a BBH, which is not as much a cone as an oddly shaped cylinder. As the BBH evolves, the shape of the loss-cone changes: (1) if the eccentricity increases significantly (but the semi-major axis remains constant) then the asymmetry in the loss-cone's shape decreases, and the co-rotating orbits are no longer more likely to be scattered. This type of loss-cone would still produce a `hole' at the centre, which will be seen as a dip in the $\sigma$ map, but no inner counter-rotating region will exist. (2) If the semi-major axis decreases significantly (but eccentricity remains low to moderate) the loss-cone would simply shrink, and the kinematic feature will only be present at very small scales. (3) Coalescence of the two BHs is the extreme case, where the loss-cone is both symmetric and minimal in size.

In our simulations, in the equal mass case, the signature forms at about the same time as the bound binary, and persists almost unchanged for the duration of the simulation (equivalent to $\sim 5\times 10^7$ years) and is likely to persist for much longer (see stability checks by \citealt{Meiron2010}). This is of course in the absence of a refilling mechanism, which our simulation lacks. The unequal mass case is very different: while the signature also forms together with the bound binary, it changes on much shorter timescales. At $t=20\,000$ (this is $10\,000$ time units after Fig. \ref{fig:KMaps-unequal}, or about ten million years) the inner counter-rotating region is all but gone. Unlike the equal mass case, it is not just `hiding' behind a more strongly co-rotating region. The dip in $\sigma$, however, becomes slightly deeper. This is curious, since as noted above, there is no relaxation mechanism in our simulation to erase the signature; if there were, the counter-rotating region would indeed be lost since eccentricity at the later snapshot is already very close to 1. It is very likely that those counter-rotating stars contribute to a secular process by which the BBH eccentricity grows, and this is why they disappear despite the lack of a relaxation mechanism. This will be looked further into in our future work.

Features such as those discussed above are on larger scales than the associated stellar clusters of the two BHs and can possibly be spatially resolved even when the BHs themselves cannot, as expected even in the nearest galaxies \citep{Yu2002}. The detection of these features may indicate the presence of a BBH currently, or one relaxation time ago, beyond which the kinematic signature is erased. However, no observation as detailed has been made yet to our knowledge. Possible candidates are the core regions of luminous elliptical galaxies which are kinematically decoupled from the main body of the galaxy, also called \textit{kinematically decoupled cores} (e.g. \citealt{Bender1988}).

Take for example elliptical galaxy NGC~4365 in the Virgo cluster which has such a kinematically decoupled core. Observation made by \citet{Davies2001} using the {\it SAURON} instrument (an integral field spectrograph) on the {\it William Herschel Telescope} and reanalyzed by \citet[see figure 7 there]{vandenBosch2008} show an inner structure, approximately 1 kpc wide, which is tilted by $82^\circ$ with respect to the rotation axis in outer radii. This region is tentatively identified with the outer, co-rotating region in Figs. \ref{fig:KMaps-unequal} and \ref{fig:KMaps-equal} in this work. While the physical size of the maps on the leftmost columns in those figures is $\sim$ twenty times smaller (assuming $M_\bullet=10^8\ \Msun$) than the decoupled core region in fig. 7 of \citet{vandenBosch2008}, the rotation presumably continues beyond the part which is shown (in fact, our entire simulated region is smaller than only $\sim 150$ pc). The pixel scale of the observations was 0.8 arcsec ($\sim 80$ pc) and thus too large to detect a possible counter-rotating inner region.

Qualitatively compared, the maps for $\mu$ and $h_3$ are similar in shape and value (in particular to the equal mass case). Note in particular that in both simulation and observational data the signs of $\mu$ and $h_3$ are opposite; which means that the retrograde wing of the line-of-sight velocity distribution is wider and the prograde wing is steeper. This feature was also discussed by \citet{mm01} and noted in observations by \citet{Bender1994}. In our simulations, this region is rotating solely because of the angular momentum transferred to the stars from the inspiralling binary, while in the full merger simulations of \citet{mm01}, and presumably in reality, much of the angular momentum should be due to the galaxy merger (i.e. the initial orbital angular momentum of the two bulges with respect to their centre of mass). It is thus curious then that the match in terms of value is so good.

The kinematic maps from the results of \citet{mm01} do not show the the counter-rotating signature on small scales (nor has it been noted elsewhere in the literature before \citealt{Meiron2010}). These were based on $N$-body simulations of $\sim 10^5$ particles, carried out on scales $\sim 100$ time larger than here (the large scale was required to simulate the full merger of the two bulges). As a result, there were only $\sim 10^3$ particles in their study inside the sphere of influence. Due to the implied large statistical errors in that study, it was not possible to probe the stellar kinematics on the scale of $r_{\rm infl}$ and closer to the BBH. Furthermore, the unrealistically small black hole mass to star mass ratio in that study caused large Brownian motion, which could likely erase the counter-rotating signature.

\section{Conclusions}\label{sec:conclusions}

We analyzed two numerical simulations of inspiralling BHs, focusing on the kinematic trail left by the binary in the stellar population. We discovered a specific signature in the LOSVD moments, which we present as high-resolution 2D maps that could serve as templates for integral field spectroscopy observations of nearby galactic centres. The discovery of such signatures may help census the population of supermassive black hole binaries and refine signal rate predictions for future space-based low frequency gravitational wave detectors. Our findings are summarized as follows:

\begin{enumerate}
\item There is an inner region with average rotation in the opposite direction to that of the BBH, it is surrounded by a larger outer co-rotating region. The inner region is detectable in projection via the mean of the LOSVD and/or its third GH moment.
\item The counter-rotating region is stable for the lifetime of the simulation in the equal mass case, but disappears in the unequal mass case when the BBH eccentricity approaches unity. A connection between eccentricity growth and the disappearance of this signature was suggested.
\item The inner region also shows a dip in $\sigma$; this is due to efficient evacuation of stars from the vicinity of the BBH.
\item The ejected stars leave behind a mass deficit, which was suggested to explain the core structure of massive ellipticals.
\end{enumerate}
Our future work will explore the lifetime of the signature for different initial models, as well as the eccentricity evolution in BBH systems. We will do so using both $N$-body simulations and approximate methods.

\bsp

\label{lastpage}

\end{document}